\documentclass{article}

\usepackage{bm}
\usepackage{dsfont}
\usepackage{amsmath}
\usepackage{mathtools}
\usepackage{amssymb}
\usepackage{graphicx}
\usepackage{graphics}
\usepackage[mathcal]{euscript}
\usepackage{authblk}
\usepackage[colorlinks=true,linkcolor=blue,allcolors=blue]{hyperref}
\usepackage{geometry}
\usepackage{xcolor}

\newcommand*{\QEDB}{\hfill\ensuremath{\square}}%
\newcommand*{\PROOF}{\noindent\textit{Proof: }}%

\usepackage{hyperref}

\newtheorem{assumption}{Assumption}

\newtheorem{theorem}{Theorem}
\newtheorem{proposition}{Proposition}

\newtheorem{definition}{Definition}

\newcommand\blfootnote[1]{%
  \begingroup
  \renewcommand\thefootnote{}\footnote{#1}%
  \addtocounter{footnote}{-1}%
  \endgroup
}

\title{\textbf{NSM Converges to a k-NN Regressor \\ Under Loose Lipschitz Estimates}}

\author{\normalsize Emilio T. Maddalena and Colin N. Jones}
\date{}
\begin{document}
\maketitle
\thispagestyle{empty}
\vspace{-0.5cm}
\begin{abstract}
Although it is known that having accurate Lipschitz estimates is essential for certain models to deliver good predictive performance, refining this constant in practice can be a difficult task especially when the input dimension is high. 
In this work, we shed light on the consequences of employing loose Lipschitz bounds in the Nonlinear Set Membership (NSM) framework, showing that the model converges to a nearest neighbor regressor (k-NN with $k=1$). This convergence process is moreover not uniform, and is monotonic in the univariate case. An intuitive geometrical interpretation of the result is then given and its practical implications are discussed.
\blfootnote{This work has received support from the Swiss National Science Foundation under the RISK project (Risk Aware Data-Driven Demand Response), grant number 200021 175627.}
\blfootnote{E. T. Maddalena and C. N. Jones are with \'Ecole Polytechnique F\'ed\'erale de Lausanne (EPFL), Switzerland. E-mails: \texttt{\{emilio.maddalena,colin.jones\}@epfl.ch}.} \\[10pt]
\end{abstract} \smallbreak

\section{Introduction}

Non-parametric models were once considered computationally too expensive to be employed in practice. However, due to the increasing availability of computational power and the decrease of hardware costs, these tools are increasingly being embraced by the control community. In \cite{kammer2017data} and \cite{nicoletti2019data} for instance, the authors use scattered frequency domain data to design off-line robust controllers for an atomic force microscope and dc-dc power converters, respectively. Gaussian processes (GPs) are an example of another popular non-parametric modeling technique \cite{liu2018gaussian,de2018real}. An early experimental investigation of GPs in control was reported in \cite{likar2007predictive}, where a gas-liquid separation plant was considered with a sampling period of $T_\text{samp} = 15$ s. More recent works have employed GPs to tackle the control of much faster systems such as autonomous racing cars with $T_\text{samp} = 20$ ms \cite{hewing2019cautious}, and robotic arms with $T_\text{samp} = 1$ ms \cite{carron2019data}. Besides their use in model predictive control (MPC) schemes as in the papers cited above, GPs have also been utilized to refine the parameters of classical state-feedback and proportional-integral-derivative (PID) compensators \cite{berkenkamp2016safe}.

Certainly, the use of these statistical regression tools coming from other communities has been causing a change in the way system identification is performed in controls \cite{ljung2020shift}. Moreover, there has also been an increasing concern regarding the safety of physical systems that operate in closed-loop based on the so called `learned models' -- see \cite{hewing2019learning} for a recent review on the subject. This is especially the case when elements of on-line learning are present. In order to overcome these issues, researchers are currently carrying out rigorous analysis of these modeling procedures to assess their uncertainties and design appropriate robust controllers \cite{umlauft2019feedback,beckers2019stable}. 

As opposed to the Bayesian techniques described previously, the Nonlinear Set Membership (NSM) approach to system identification \cite{milanese2004set} does not rely on the presence of priors, nor does it quantify uncertainties in a statistical fashion. This non-parametric methodology assumes a certain degree of regularity of the unknown target function $f(x)$, namely that it is Lipschitz continuous with a known upper bound. Based purely on collected data-points, \textit{tight} deterministic bounds on the function values $f(x)$ at unobserved points can be established, making it an interesting modeling alternative. The theoretical foundations of the NSM technique were recently generalized to H\"older continuous functions \cite{calliess2016lazily,blaas2019localised}, extended to a large class of problems \cite{milanese2011unified}, and also to more abstract function spaces \cite{novara2019nonlinear}. Applications of this methodology can be found in the context of vehicle yaw reference-tracking \cite{canale2008vehicle}, electrical microgrids scheduling \cite{la2019set}, and approximating general linear MPC control laws \cite{canale2009set}. In a broader sense, similar set-membership ideas were recently used to study the simulation of linear systems with guaranteed accuracy \cite{lauricella2020set}.

The quality of a prediction given by an NSM model strongly depends on how close the employed Lipschitz estimate is to the best constant, i.e., the lowest valid one. From a practical viewpoint however, refining this quantity can easily become a daunting task, especially in cases where the input dimension is large (see \cite[Sec. 4.1]{bunin2013sufficient}, and the simulation results in \cite{canale2014nonlinear}). Statistical methods exist to deal with the problem \cite{wood1996estimation}, but are limited to the univariate case. Ad hoc procedures followed by an augmentation to create a `safety margin' seem to be a rather common way of estimating it in practical scenarios. A question thus naturally arises regarding NSM models: \textit{besides enlarging the error bounds, what effects do loose Lipschitz estimates have on the regressor itself?}

Contributions: We answer the above question in this paper, showing that regardless of the input dimension, the NSM model converges to another well-known non-parametric model, namely the nearest neighbor regressor (k-NN with $k=1$). This convergence is shown under the $\mathcal{L}^2$ functional norm, and is moreover not uniform. The result is nevertheless valid for any finite-dimensional norm $||\cdot||$ chosen to define the NSM and nearest neighbor regressors. Although not monotonic in general, the discrepancy between both models is shown to only decrease as the Lipschitz estimate is loosened in the univariate case. We emphasize that it is beyond the scope of this work to measure the distance of any of these models and the unknown ground-truth, as our goal is to build a bridge between methodologies. Still, we do discuss the implications of our results and also practical ideas on how to efficiently evaluate these piecewise functions.

\section{Preliminaries}

\textit{Notation}: Given a set $S$, $\text{int}(S)$ represents its interior. The sets $X \subset \mathbb{R}^{n_x}$ and $Y \subset \mathbb{R}$ denote compact subsets of Euclidean spaces. $[N]$ denotes the set of integers $\{1,\dots,N\}$. $||x||$ represents any $\ell_p$ norm, whereas $||f||_2 := (\int_{X} f(x)^2 dx)^{\frac{1}{2}}$ and $||f||_\infty := \max_{x \in X} |f(x)|$ are respectively the usual $\mathcal{L}^2$ and $\mathcal{L}^\infty$ function norms. Herein we will say a finite collection of sets $\mathtt{X}^n$, $n \in [N]$ , is a \textit{partition} of $X$ if: $\cup_{n=1}^{N} \mathtt{X}^n = X$ and $\text{int}(\mathtt{X}^i) \cap \text{int}(\mathtt{X}^j) = \emptyset$, $\forall i,j \in [N]$. A function $f:X \rightarrow Y$ is said to be Lipschitz continuous if $\exists L > 0 : \forall x_1, x_2 \in X, \, |f(x_1)-f(x_2)| \leq L ||x_1-x_2||$. The lowest such constant is known as the best Lipschitz constant of $f$, denoted as $L^\star$. We assume $L^\star$ is unknown, but upper bounds $L \geq L^\star$ are known.
\bigbreak

\noindent \textit{The dataset:} Consider a collection of labeled samples 
\begin{equation}
    D = \{x_n,y_n\}_{n\in [N]}
    \label{eq.dataset}
\end{equation}
where the following input-output relation holds $y_n = f(x_n)$, $\forall n = 1, \dots, N$ and the function $f:X \rightarrow Y$ is Lipschitz continuous with $L$ as a valid constant.

\begin{assumption}
\label{assu.images}
Not all sample images are equal, i.e., $\exists n,i \in [N]: y_n \neq y_i$.
\end{assumption}

The assumption above is necessary for the convergence problem to be non-trivial. Indeed, if all points in $D$ have the same image, it is straightforward to show that there is no discrepancy between the two regressors regardless of $L$.
\bigbreak

\noindent \textit{The k-NN regressor:} Given a dataset $D$, a measure of distance induced by $||\cdot||$ on $X$, and a positive integer $k\leq N$, the k-NN regressor is then defined as\footnote{Alternative methods of assigning distinct weights to the neighbors exist, favoring the nearest ones for instance. Averaging is the simplest form of weight assignment.}
\begin{equation}
    \label{eq.kNNreg}
    f_{\text{NN}}(x) \coloneqq \frac{1}{k} \sum_{n \in \mathcal{N}(x)} y_n
\end{equation}
where $\mathcal{N}(x)$ is the index set of the k-nearest neighbors of $x$, i.e., points $x_i$ that satisfy $||x-x_i|| \leq ||x-x_n||, \; \forall i \in \mathcal{N}(x), \forall n \in [N] \backslash \mathcal{N}(x)$. In our rather particular case, with $k=1$, $f_{\text{NN}}(x)$ simply replicates at $x$ the image of its closest data-point in $D$. 

\begin{assumption}
In case a point $x \in X$ has multiple nearest neighbors, we assume a selection rule exists, causing $\mathcal{N}(x)$ to be always a singleton.
\end{assumption}

The previous assumption was needed to define a proper nearest neighbors regressor without ambiguity. The selection rule does not impact our analysis since it only affects a subset of the domain with measure zero.

A Voronoi cell associated with a single point $x_n \in D$ is defined as
\begin{equation}
    \mathtt{C}^n \coloneqq \{x \in X \, | \, ||x-x_n|| \leq ||x-x_j||, \forall j \in [N]\}
\end{equation}
and the Voronoi diagram (VD) of a dataset $D$, as the partition of the domain $X$ induced by the Voronoi cells $\mathtt{C}^1,\dots,\mathtt{C}^N$. The nearest neighbor (NN) regressor then assumes the simplified form
\begin{equation}
    f_{\text{NN}}(x) = y_n, \; \text{if~} x \in \mathtt{C}^n, \, \exists n \in [N]
    \label{eq.NN}
\end{equation}
derived from \eqref{eq.kNNreg} with $k=1$.
\bigbreak

\noindent \textit{The NSM regressor:} We consider noiseless measurements as already indicated in \eqref{eq.dataset} to simplify our presentation. The derivations can be repeated even when the dataset $D$ is affected by an unknown but bounded noise.

Given the dataset $D$ and a Lipschitz constant estimate $L$ for the unknown ground-truth, define the auxiliary `ceiling' and `floor' functions respectively as
\begin{align}
    & \overline{f}(x) \coloneqq \min_{n \in [N]} y_n + L ||x-x_n|| \\
    & \underline{f}(x) \coloneqq \max_{n \in [N]} y_n - L ||x-x_n|| 
\end{align}
The nonlinear set membership regressor is then defined as 
\begin{equation}
    f_{\text{NSM}}(x) \coloneqq \frac{1}{2} \; \big( \, \overline{f}(x) + \underline{f}(x)\big)
\end{equation}
As in the previous case, it is convenient to define a domain partition over which $\overline{f}$ and $\underline{f}$ enjoy a simpler representation. 

Additively weighted Voronoi diagrams (AVD), also known as hyperbolic Voronoi diagrams, are an extension of VD where the n-th cell is defined as
\begin{equation}
    \{x \in X \, | \, ||x-x_n|| \leq ||x-x_j|| + \eta_{n,j} \,, \forall j \in [N]\}
    \label{eq.AVD}
\end{equation}
where $\eta_{n,j}$ can be regarded as a bias. Next, construct two AVDs with cells $\overline{\mathtt{C}}^n$ and $\underline{\mathtt{C}}^n$ as in \eqref{eq.AVD}, respectively with $\overline{\eta}_{n,j} = (y_j - y_n)/L$ and $\underline{\eta}_{n,j} = (y_n - y_j)/L$. This causes the AVD domain partition to be influenced not only by the locations $x_n$, but also by the attained values $y_n$. As shown in \cite{milanese2004set}, the ceiling and floor functions can be rewritten as
\begin{align}
    & \overline{f}(x) \coloneqq y_n + L ||x-x_n||, \; \text{if~} x \in \overline{\mathtt{C}}^n, \, \exists n \in [N] \label{eq.ceiling} \\
    & \underline{f}(x) \coloneqq y_n - L ||x-x_n||, \; \text{if~} x \in \underline{\mathtt{C}}^n, \, \exists n \in [N] \label{eq.floor}
\end{align}
describing two piecewise conic functions defined over two distinct AVD. As a result, $f_{\text{NSM}}$ is also defined piecewise, on the intersection of the AVDs, being in general piecewise non-linear if $||\cdot||$ is not the $\ell_1$ nor the $\ell_\infty$ norm.

\section{Theoretical results}

\subsection{Main findings}

We begin by introducing the auxiliary $\mathtt{A}^n$ and $\mathtt{B}^{n,m}$ cells that define a new partition of the domain $X$.

\begin{definition} 
\label{def:AandB}
Let $\mathtt{A}^{n,m} \coloneqq (\overline{\mathtt{C}}^n \cap \underline{\mathtt{C}}^m), n,m \in [N], \, m \neq n$, and $\mathtt{B}^n \coloneqq (\overline{\mathtt{C}}^n \cap \underline{\mathtt{C}}^n), n \in [N]$.
\end{definition}

\begin{proposition}
\label{prop.B}
For any $n \in [N]$, $\mathtt{B}^n \neq \emptyset$. Moreover, let $x \in \mathtt{B}^n$, then $x \in \mathtt{C}^n$.
\end{proposition}
\PROOF
From the definition of $\mathtt{B}^n$ we have
\begin{equation}
    \label{eq.Baux}
    \mathtt{B}^{n} = \left\{x \in X \; \left| \,
    \begin{aligned}
        & ||x-x_n|| \leq ||x-x_i|| + \frac{y_j-y_n}{L}, \forall j \in [N] \\ 
        & ||x-x_n|| \leq ||x-x_j|| + \frac{y_n-y_j}{L}, \forall j \in [N]
    \end{aligned} 
    \right.
    \right\}
\end{equation}
Since for all $j \in [N]$ either $y_j-y_n \leq 0$ or $y_n-y_j \leq 0$, then $||x-x_n|| - ||x-x_j|| \leq 0 \implies x \in \mathtt{C}^n$. Consider the case $x = x_n$. If $y_j - y_n \geq 0$, then the first inequality in \eqref{eq.Baux} holds due to the norm being a positive function, whereas the second holds since the dataset $D$ was generated by a Lipschitz continuous function with $L$ as a valid constant. If on the other hand $y_j - y_n \leq 0$, the arguments apply in the reverse order. Hence, the sets $\mathtt{B}^1,\dots,\mathtt{B}^N$ will always respectively contain at least $x_1,\dots,x_N.\QEDB$ 
\bigbreak

The domain was partitioned into three different diagrams: a VD, a ceiling AVD, and a floor AVD. Each point in the domain thus belongs to a cell in each one of them. Given the cells to which a specific $x$ belongs, we establish in the following proposition an order for their associated images.

\begin{proposition}
\label{prop.order}
If $x \in \mathtt{A}^{n,m}$ and also $x \in \mathtt{C}^{p}$, with $p \neq n$ and $p \neq m$, then $y_n \leq y_p \leq y_m$ must hold.
\end{proposition}
\PROOF 
Follows from the feasibility of the set of inequalities in $\overline{\mathtt{C}}^n \cap \underline{\mathtt{C}}^m \cap \mathtt{C}^p. \QEDB$

\begin{proposition}
\label{prop.bounds}
Let $F_\text{NSM} := \max_{n \in [N]} |y_n|$ be the maximum over the dataset labels absolute values, then $F_\text{NSM} \geq |f_\text{NSM}(x)|, \forall x \in X$, regardless of its Lipschitz estimate. 
\end{proposition}

\PROOF
The analysis is done region by region since there is only a finite number of them. From \eqref{eq.ceiling} and \eqref{eq.floor}, inside the $\mathtt{B}^n$ cells, $f_\text{NSM}(x)$ always attains $y_n$. Assume now $\exists x \in \mathtt{A}^{n,m}: f_\text{NSM}(x) > y_m$. This is equivalent to $\frac{1}{2} (y_n + y_m + L(||x-x_n|| - ||x-x_m||)) > y_m$, which implies $||x-x_n|| > ||x - x_m|| + \frac{y_m-y_n}{L} \implies x \not\in \overline{\mathtt{C}}^n \implies x \not\in \mathtt{A}^{n,m}$, leading to a contradiction. Therefore, $\forall x \in \mathtt{A}^{n,m}, \, f_\text{NSM}(x) \leq y_m$. An analogous contradiction can be constructed to show that $\forall x \in \mathtt{A}^{n,m}, \, f_\text{NSM}(x) \geq y_n$. Hence, on each $\mathtt{A}^{n,m}, \, f_\text{NSM}(x)$ is bounded by $\max(|y_n|,|y_m|)$. On the whole domain, the regressor absolute value is thus tightly bounded by $F_\text{NSM}. \QEDB$
\bigbreak

In what follows, $L_l,L_{l+1},\dots$ denotes a sequence of strictly increasing Lipschitz constants indexed by $l$, which give rise to a sequence of NSM regressors $f_{\text{NSM},l},f_{\text{NSM},l+1},\dots$. The error function is then defined as
\begin{equation}
    \label{eq.error}
    e_l \coloneqq f_\text{NN} - f_{\text{NSM},l}
\end{equation}
with a point-wise subtraction.

To simplify our notation, the dependence of the functions and sets on the sequence index $l$ will be omitted, and made explicit only when convenient. The main convergence theorem is stated next, and its proof relies on bounding the worst-case error between the NSM and NN functions, as well as on the strict expansion of certain $\mathtt{B}^n$ sets.

\begin{theorem}
\label{th:convergence}
Under the $||\cdot||_2$ functional, $f_{\text{NSM}} \rightarrow f_{\text{NN}}$ as $l \rightarrow \infty$ with $k=1$ neighbor.
\end{theorem}

\PROOF
The aim is to show that $\forall \epsilon > 0$, $\exists l$ with its associated $L_l$ s.t. $||f_{\text{NN}} - f_{\text{NSM}}||_2 < \epsilon$ for any equal or larger constant $L_l,L_{l+1},\dots$.

From \eqref{eq.ceiling} and \eqref{eq.floor}, begin by writing $f_{\text{NSM}}$ as
\begin{equation}
    f_{\text{NSM}}(x) = 
     \begin{cases}
       y_n, \ \, \text{if~} x \in \mathtt{B}^n, \, \exists n \in [N]\\
       \frac{y_n+y_m}{2} + \frac{L}{2}(||x-x_n||-||x-x_m||), \\
       \qquad \text{if~} x \in \mathtt{A}^{n,m}, \, \exists n,m \in [N]
     \end{cases}
\end{equation}

From \eqref{eq.error} and Proposition~\ref{prop.B}, the error becomes

\begin{equation}
    \label{eq:error}
    e_l(x) = 
     \begin{cases}
       0, \quad \text{if~} x \in \mathtt{B}^n, \, \exists n \in [N] \\
       y_p - \frac{y_n+y_m}{2} - \frac{L_l}{2}(||x-x_n||-||x-x_m||), \\
       \qquad \text{if~} x \in \mathtt{A}^{n,m}, \, \exists n,m \in [N] \\
       \qquad \text{and} \; p = \mathcal{N}(x) \\
     \end{cases}
\end{equation}
where $\mathcal{N}(x)$ is the index of the nearest neighbor cell $\mathtt{C}^p$.

From \eqref{eq:error}, it is clear that the union of all $\mathtt{A}^{n,m}$ also contains the support\footnote{The subset of the domain where a function attains non-zero values.} of $e_l(x)$. From \eqref{eq.Baux}, the dependence of $\mathtt{B}^n$ on the Lipschitz estimate $L_l$ is
\begin{equation}
    \mathtt{B}_{l}^{n} = \left\{x \in X \, \left| 
        ||x-x_n|| \leq ||x-x_i|| + \frac{|y_i-y_n|}{L_l}, \forall i \in [N] 
    \right.
    \right\}
\end{equation}
which shows that for every $i$ the inequality is restricted by the bias if $y_i \neq y_n$, and that this restriction depends inversely on $L_l$. Due to Assumption~\ref{assu.images}, at least one $\mathtt{B}^n_l$ cell will be restricted by the bias; therefore, for that particular cell, $\mathtt{B}^n_l \subset \mathtt{B}^n_{l+1} \subset \dots$ since $L_l < L_{l+1} < \dots$. This implies that
\begin{equation}
    \bigcup_{n} \mathtt{B}_l^n \, \subset \, \bigcup_{n} \mathtt{B}_{l+1}^n \, \subset \, \dots
\end{equation}
which in turn implies 
\begin{equation}
    \label{eq.Asets}
    \bigcup_{n,m} \mathtt{A}_l^{n,m} \, \supset \, \bigcup_{n,m} \mathtt{A}_{l+1}^{n,m} \, \supset \, \dots
\end{equation}
as the union over $n$ and $m$ of all $\mathtt{B}_l^{n}$ and $\mathtt{A}_l^{n,m}$ cells forms a partition of the domain $X$.

Given that $f_{\text{NN}}$ in \eqref{eq.NN} is bounded and has only a finite number of discontinuities, which have measure zero, $f_{\text{NN}}$ is integrable. In view of this fact and the continuity of $f_{\text{NSM}}$, $e_l$ is integrable $\forall l$ and its norm $||e_l||_2$ is defined. 
Let $F_{\text{NSM}} > 0$ and $F_{\text{NN}} > 0$ be constants such that $F_{\text{NSM}} > |f_{\text{NSM}}(x)|$ and $F_{\text{NN}} > |f_{\text{NN}}(x)|, \forall x \in X$ (the former was defined in Proposition~\ref{prop.bounds}). The inequality $F_{\text{NSM}} + F_{\text{NN}} =: F_{\text{e}} > |e_l(x)|$ then holds $\forall x, \, \forall l$. 

Now define the auxiliary constant $\gamma_l^{n,m} \coloneqq \int_{\mathtt{A}_l^{n,m}}1 \, dx$, which is guaranteed to exist and to be finite due to the sets $\mathtt{A}_l^{n,m}$ being compact, and $\sigma_l \coloneqq \sum_{m,n} \gamma_l^{n,m}$. We can then upper-bound the square of a given error norm by
\begin{subequations}
\label{eq.largeSeq}
\begin{align}
||e_l||^2_2 & = \int_X e_l^{\, 2}(x) dx \\
    & = \sum_{n,m} \int_{\mathtt{A}_l^{n,m}} e_l^{\, 2}(x) \, dx + \sum_{n}  \int_{\mathtt{B}_l^{n}} e_l^{\, 2}(x) \, dx \\
    & =\sum_{n,m}  \int_{\mathtt{A}_l^{n,m}} e_l^{\, 2}(x) \, dx + 0 \\
    & < F_e^{\, 2} \sum_{n,m} \gamma_l^{n,m} \\
    & = F_e^{\, 2} \, \sigma_l
\end{align}
\end{subequations}
Moreover, $||e_l||^2_2 < F_e^{\, 2} \sigma_l \implies ||e_l||_2 < F_e \sqrt{\sigma_l}$. As the union of the sets in \eqref{eq.Asets} are strictly decreasing, then so is their measure, leading to $\lim_{l \rightarrow \infty} \sigma_l = 0$ monotonically. Due tho the later fact and $||e_l||_2$ being bounded from below by zero, it follows from the squeeze theorem that $\lim_{l \rightarrow \infty} ||e_l||_2 = 0$ and therefore $f_{\text{NSM}} \xrightarrow{l \rightarrow \infty} f_{\text{NN}}. \QEDB$ 

\begin{theorem}
\label{thm.nonUniform}
The converge in Theorem~\ref{th:convergence} is not uniform, i.e., $f_{\text{NSM}} \nrightarrow f_{\text{NN}}$ as $l \rightarrow \infty$ with respect to $||\cdot||_\infty$.
\end{theorem}

\PROOF
It suffices to show that $||e_l||_\infty = ||e_{l+1}||_\infty, \forall l$. Consider the error expression in \eqref{eq:error} with $x \in \mathtt{A}^{n,m}$ and $x \in \mathtt{C}^p$. We analyze the case when $y_p > \frac{y_n+y_m}{2}$, whereas $y_p < \frac{y_n+y_m}{2}$ follows by analogy. Let $y_p > \frac{y_n+y_m}{2}$ and $||x-x_n|| \geq ||x-x_m||$, then $|e_l(x)|$ is maximized when $||x-x_n|| =||x-x_m||$, leading to $|e_l(x)| = |y_p - \frac{y_n+y_m}{2}|$. On the other hand, if $||x-x_m|| \geq ||x-x_n||$, then $|e_l(x)|$ is maximized when $||x-x_m|| =||x-x_n|| + \frac{y_m-y_n}{L_l}$, leading to $|e_l(x)| = |y_p - y_n|$. The previous analysis is valid for any region of $e_l(x)$. Since all local maxima are independent of the Lipschitz estimate $L_l$ and $||e_l||_\infty$ is the maximum over all of them, we conclude that $||e_l||_\infty = ||e_{l+1}||_\infty, \forall l. \QEDB$ 

\subsection{Additional analysis}

In this section we inspect parts of the domain where the absolute error is guaranteed to decrease monotonically with the index $l$. This will be expressed in terms of two sufficient conditions later used to establish an additional result.

\begin{proposition}
\label{prop.errorDec1}
Let $x \in \mathtt{A}_l^{n,m}$, $x \in \text{int}(\mathtt{C}^n)$, and also $x \in \mathtt{A}_{l+1}^{n,m}$ at the next index. In this case, $|e_l(x)| > |e_{l+1}(x)|$. The same conclusion holds if $x \in \mathtt{A}_l^{n,m}$, $x \in \text{int}(\mathtt{C}^m)$, and $x \in \mathtt{A}_{l+1}^{n,m}$.
\end{proposition}

\PROOF
Consider $x \in \mathtt{A}_l^{n,m}$, $x \in \text{int}(\mathtt{C}^n)$ and $x \in \mathtt{A}_{l+1}^{n,m}$. In view of this and of Proposition~\ref{prop.order}, the error in \eqref{eq:error} becomes 
\begin{equation}
    e_l(x) = \underbrace{\frac{y_n-y_m}{2}}_{=:a \, < 0} -
    \underbrace{\frac{L_l}{2}(||x-x_n||-||x-x_m||)}_{=: b \, < 0}
\end{equation}
Note moreover that $|a| \geq |b|, \forall l,$ since $|b| > |a| \implies x \not\in \mathtt{A}^{n,m}$ (see the proof of Proposition~\ref{prop.bounds}). At the next index sequence, $l+1$, the $a$ stays constant, but $b$ decreases as $L_{l+1} > L_l$. The absolute error has therefore decreased, i.e., $|e_l(x)| > |e_{l+1}(x)|$. Analogous arguments can be made if $x \in \text{int}(\mathtt{C}^m). \QEDB$
\bigbreak

As the Lipschitz estimate changes, the AVD cells are affected and the partitions change. According to the following proposition, given a point in the domain, if the pair of AVD cells that contains it changes from index $l$ to $l+1$, then the difference between images associated these cells must have decreased. 

\begin{proposition}
\label{prop.errorDec2}
Let $x \in \mathtt{A}_l^{n,m}$ at step $l$ and $x \in \mathtt{A}_{l+1}^{o,m}$ at step $l+1$, $o \neq n$, then necessarily $y_o \geq y_n$. Similarly, if $x \in \mathtt{A}_l^{n,m}$ and $x \in \mathtt{A}_{l+1}^{n,o}$, $o \neq m$, then necessarily $y_o \leq y_m$. If, more specifically, $x \in \mathtt{A}_l^{n,m}$ and $x \in \mathtt{B}_{l+1}^{n}$ or $x \in \mathtt{B}_{l+1}^{m}$, then $|e_l(x)| > |e_{l+1}(x)|$.
\end{proposition}

\PROOF
The first case is shown, whereas the second follows by analogy. Given $x \in \mathtt{A}^{n,m}$, $x \in \mathtt{A}_{l+1}^{o,m}$ and the definitions of the AVD cells in \eqref{eq.AVD}, we have that
\begin{equation}
    \begin{aligned}
        & ||x-x_n|| \leq ||x-x_o|| + \frac{y_o-y_n}{L_l} \\ 
        & ||x-x_o|| \leq ||x-x_n|| + \frac{y_n-y_o}{L_{l+1}}
    \end{aligned}
\end{equation}
leading to $||x-x_n|| - \frac{y_o-y_n}{L_l} \leq ||x-x_o|| \leq ||x-x_n|| + \frac{y_n-y_o}{L_{l+1}}$, which is only feasible if $y_o \geq y_n$.

The strict error decrease is implied by $|e_l(x)| > 0, \forall x \in \text{int}(\mathtt{A}_l^{n,m})$, and $|e_{l+1}(x)| = 0, \forall x \in \mathtt{B}_{l+1}^{n}$ and $\forall x \in \mathtt{B}_{l+1}^{m}. \QEDB$
\bigbreak

Even though Proposition~\ref{prop.errorDec1} and Proposition~\ref{prop.errorDec2} present local monotonic error decrease results, there might be areas of the domain where $|e_l(x)|$ can increase. Whether or not the $\mathcal{L}_2$ norm of the function will strictly decrease at every new index depends on its integral over the whole domain. Nevertheless, in the particular case of a univariate model, an additional property can be established.

\begin{theorem}
If the model is univariate, i.e., $X \subset \mathbb{R}$, then $f_\text{NSM} \rightarrow f_\text{NN}$ as $l \rightarrow \infty$ with respect to $||\cdot||_2$ monotonically.
\end{theorem}

\PROOF
Begin by sorting the dataset $D$ in an increasing order of features $x_1 \leq x_2 \leq \dots \leq x_N$. Consider two cases:

\textit{Case 1}: If $\exists n, m \in [N]$ such that $x$ lies in between them, $x_n \leq x \leq x_m$, then clearly $x \in \mathtt{C}^n$ or $x \in \mathtt{C}^m$. Let $y_n \leq y_m$. Suppose for the sake of building a contradiction that $x \in \overline{\mathtt{C}}^p$. If $x$ is greater than $x_p$ and $x_n$, then $|x-x_p| - |x-x_n| = |x_n-x_p|$ and $x \in \overline{\mathtt{C}}^p \implies |x-x_p| \leq |x-x_n| + \frac{y_n-y_p}{L} \implies |y_n-x_p| \geq L |x_n-y_p| \implies$ $L$ is not a valid Lipschitz constant, establishing a contradiction. Similar arguments can be made for $x_p \geq x_m$, and in the case $x \in \underline{\mathtt{C}}^p$. Thus, $x$ can only be contained in the AVD cells $\overline{\mathtt{C}}^n$ and $\underline{\mathtt{C}}^m$.

\textit{Case 2}: If $x \leq x_1$, then $x \in \mathtt{C}^1$. Moreover, $x \in (\overline{\mathtt{C}}^1 \cap \underline{\mathtt{C}}^1)$ since belonging to any other AVD cell would invalidate $L$, leading to a contradiction. Analogous arguments hold if $x \geq x_N$. Therefore, $x$ can only be contained in AVD cells with the same index of its VD cell.

Due to the previous facts, the VD and AVD cells to which any point $x \in X$ belongs must be the ones associated with either the two feature points $x_n$ and $x_m$ that surround it (\textit{Case~1}), or the smallest/largest point $x_1$/$x_N$ (\textit{Case~2}). Hence, if $x \in \mathtt{A}_{l}^{n,m}$, then $x \in \mathtt{A}_{l+1}^{n,m}$ or $x \in \mathtt{B}_{l+1}^{n}$ or $x \in \mathtt{B}_{l+1}^{m}$. In any of the three cases, by Proposition~\ref{prop.errorDec1} and Proposition~\ref{prop.errorDec2}, $|e_{l}(x)| \geq |e_{l+1}(x)|$. If, on the other hand, $x \in \mathtt{B}_{l}^n$, $x$ will also belong to $\mathtt{B}_{l+1}^n$ and $e_l(x) = e_{l+1}(x)$. Therefore, for any $x \in X$, $|e_{l}(x)| \geq |e_{l+1}(x)| \implies e_{l}^{\, 2}(x) \geq e^{\, 2}_{l+1}(x) \implies$ the convergence is monotonic.$\QEDB$

\section{Discussion}

Refining Lipschitz estimates of an unknown function is not straightforward in the NSM framework, where the ground-truth $f$ and clearly also its gradient $\nabla f$ are assumed to be unknown. This is refered to as the `black-box' case, as opposed to scenarios where either $f$ or $\nabla f$ are known \cite{wood1996estimation}. Methods available in the literature usually prove convergence $L \rightarrow L^\star$ from below, thus yielding lower bounds instead of upper bounds (see for instance \cite{calliess2016lazily,malherbe2017global,blaas2018scalable}). Other approaches propose the \textit{optimization} of $L$ as a decision variable \cite{calliess2017lipschitz}. A statistical procedure for univariate models was proposed in \cite{wood1996estimation}; nevertheless, care should be taken before employing it in the NSM scenario since it requires the dataset to be composed of independent samples. In classical system identification however, samples are usually highly dependent on each other, as they are the different time-instants of a trajectory. In their original paper \cite{milanese2004set}, the authors propose partitioning the dataset into two parts: one to compute the `$\gamma$ surface' of non-falsified constants and error bounds, and another to construct the NSM model. As pointed out by the authors however, parameters chosen from the surface might be invalidated by future data, meaning that there is no guarantee that the employed constant is indeed a valid Lipschitz constant for the underlying ground-truth.

As it is difficult to refine the Lipschitz hyperparameter $L$ over time staying always above $L^\star$, it is common in practice to apply safety factors to ensure $L \geq L^\star$. The consequences of having loose estimates should therefore be kept in mind when using the resulting regressor. More specifically, it is well known that nearest neighbor models present strong variance \cite{altman1992introduction}, having zero error in the training phase, but severely overfitting the data, thus causing high test errors. Since the NSM model tends to a discontinuous function with constant plateau regions, numerical problems might arise when employing it in numerical optimization procedures such as MPC. With the goal of remedying this issue, smooth surrogates were recently proposed in \cite{manzano2019output} and \cite{maddalena2019}. The lack of robustness of NN models highlighted in works such as \cite{wang2017analyzing} is however not an issue, as the NSM technique is only employed in regression tasks rather than classification ones.

When the number of data-points is large, the seemingly simple task of finding the ceiling and floor AVD cells that contain a given point $x$ can become prohibitive, especially in real-time applications. Checking the containment of $x$ cell by cell has query-time complexity $\mathcal{O}(N)$. For standard Voronoi diagrams, certain $\epsilon$-approximate nearest neighbors (ANN) methods have logarithmic complexity instead \cite{arya1998optimal}. After reviewing the literature, we found that the data structure proposed in \cite{har2015approximating} -- which appears to be unknown to NSM users -- could considerably speed up NSM evaluations as it also has logarithmic time-complexity and covers distance functions with additive offsets.

\section{Numerical example}

\begin{figure*}[t!]
    \begin{center}
    \vspace{0.5cm}
    \includegraphics[width=1\textwidth]{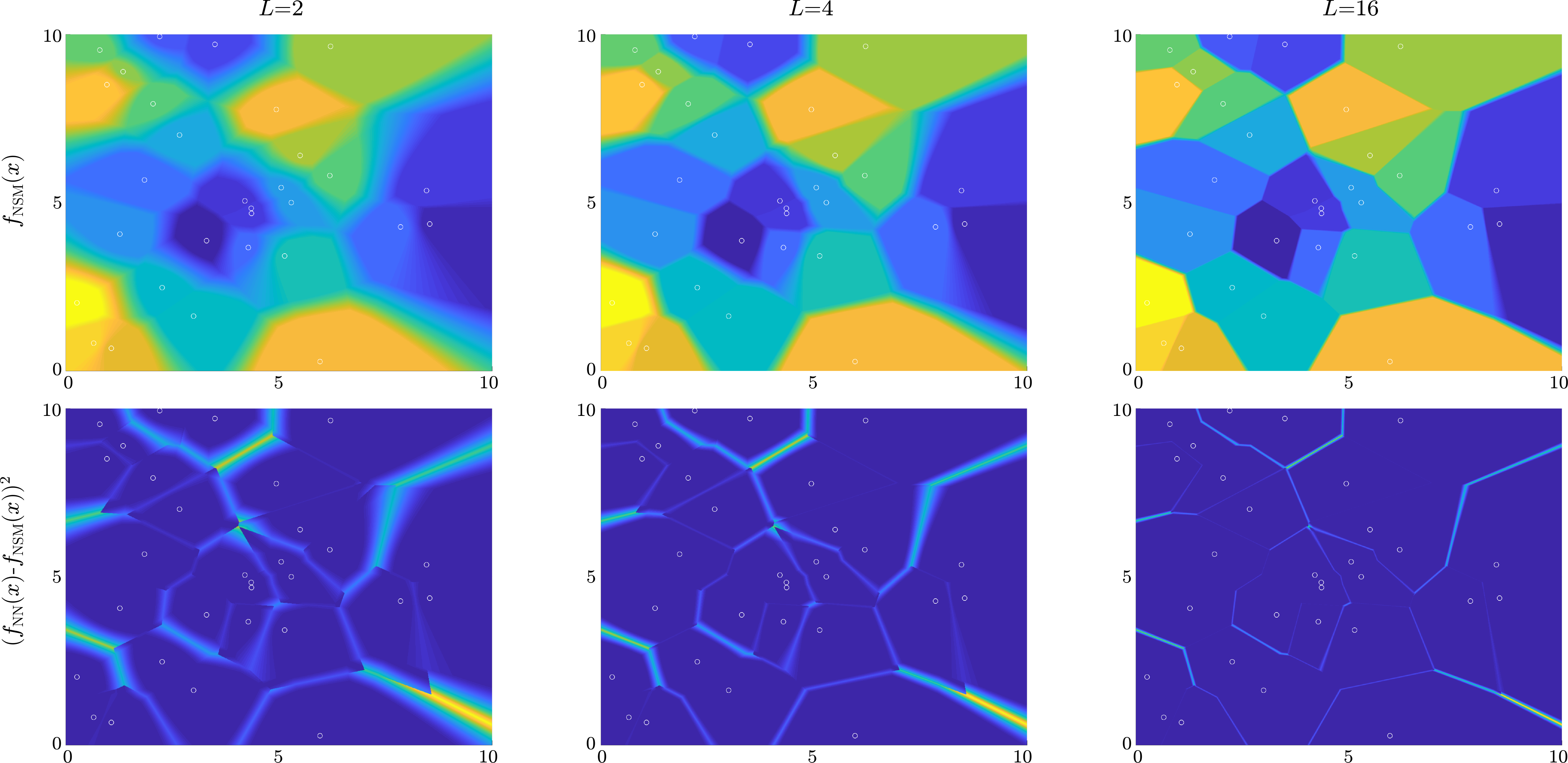}
    \caption{(Top row, from left to right) NSM regressors $f_\text{NSM}$ constructed with three different Lipschitz estimates: $L = 2$, $4$ and $16$. (Bottom row) Squared-difference $(f_\text{NN}-f_\text{NSM})^2$ between the above NSM regressor and a nearest neighbor one. Points belonging to the dataset $D$ are depicted as white circles. Low values are indicated in blue, and high values, in yellow.} 
    \label{fig.1}
    \end{center}
\end{figure*}

The ground-truth $f(x) = \cos(x_1) + \sin(x_2)$, $x = \begin{bmatrix}x_1 & x_2 \end{bmatrix}^T$ was chosen for this example and $N = 30$ samples were collected employing a uniform distribution over the domain $X = \{x \in \mathbb{R}^2 | \begin{bmatrix}0 & 0\end{bmatrix} \leq x \leq \begin{bmatrix}10 & 10\end{bmatrix}\}$. Since the domain is compact and convex, the best Lipschitz constant of $f(x)$ is the maximum value of $||\nabla f(x)||$, which was estimated with a fine grid to be $L^\star \approx 1.4142$ using the $\ell_2$ norm. NSM regressors were then computed based on three Lipschitz estimates, $L = 2, 4$, and $16$, and are shown in the top row of Figure~\ref{fig.1}. Discrepancy plots between the NSM models and a fixed nearest neighbor regressor, $(f_\text{NN}-f_\text{NSM})^2$, are instead presented in the bottom row of Figure~\ref{fig.1}.

\begin{figure}[t!]
    \begin{center}
    \includegraphics[scale=0.35]{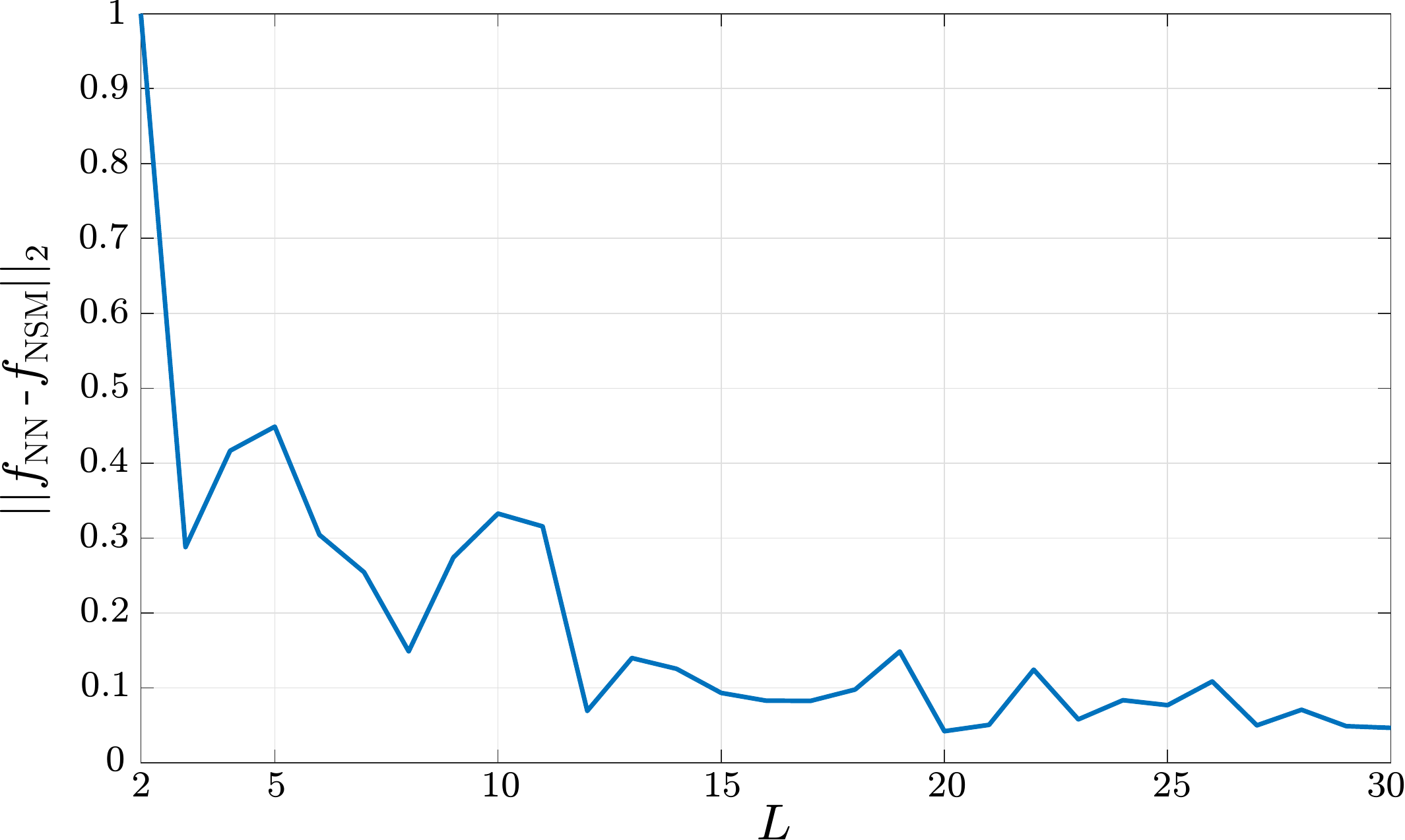}
    \caption{Normalized norm values $||f_\text{NN}-f_\text{NSM}||_2$ as a function of the Lipschitz estimate employed in the NSM regressor.} 
    \label{fig.2}
    \end{center}
\end{figure}

It is possible to see in the upper plots of Figure~\ref{fig.1} constant `plateau' regions around each sample, that is, the $\mathtt{B}^n$ cells. Those are interfaced by transition areas in which there are gradients of colors, the $\mathtt{A}^{n,m}$ cells. We moreover note that the blurred regions that separate the data-points become sharper as $L$ increases, closely resembling a standard Voronoi partition when $L=16$. In the lower plots, the zero-error $\mathtt{B}^n$ areas can be again identified around each sample with $\mathtt{A}^{n,m}$ cells interfacing them. As $L$ increases, some regions where $(f_\text{NN}-f_\text{NSM})^2 \neq 0$ shrink, preserving however their maximum values. The error $\mathcal{L}_2$ norm $||f_\text{NN}-f_\text{NSM}||_2$ was then estimated by gridding the domain with a fine mesh, and the resulting values for Lipschitz constants ranging from $2$ to $30$ are shown in Figure~\ref{fig.2}. As can be seen in the normalized plot, the discrepancy between models was reduced as the Lipschitz estimate becomes looser, but this process was not monotonic in this particular example.

\section{Conclusion}

Loose Lipschitz estimates have the effect of causing the NSM regressor to converge to a nearest neighbor model. The later is known to exhibit a strong local behavior and to possess high variance error. The looser the $L$ parameter, the closer the two additively weighted Voronoi diagrams associated with the NSM will be to the NN Voronoi partition. More precisely, the union of the $\mathtt{B}^n$ cells -- which are a subset of the VD $\mathtt{C}^n$ cells and where both regressors attain the same values -- is expanded. This convergence is not uniform in the domain, as the maximum error value is invariant to Lipschitz constant modifications. Monotonicity is present in the univariate case, but not necessarily in higher dimensions.

\bibliographystyle{IEEEtran}
\bibliography{ifacconf} 

\end{document}